\documentclass[conference]{IEEEtran}

\usepackage{yfonts} 

\usepackage{caption}
\usepackage{subcaption}
\usepackage[normalem]{ulem}
\usepackage{changepage}
\usepackage{cite}
\usepackage[utf8]{inputenc}
\usepackage{xspace}
\usepackage{balance}
\usepackage{amssymb}
\usepackage{amsmath}
\usepackage{amsthm}
\usepackage{mathtools}
\usepackage{dsfont} 
\usepackage{thmtools}  
\usepackage{cancel}
\usepackage{graphicx}
\usepackage{wrapfig}
\PassOptionsToPackage{hyphens}{url}
\usepackage[hidelinks]{hyperref}
\usepackage[hyphens]{url}
\usepackage{booktabs}
\usepackage{bm}
\usepackage{enumitem}
\usepackage[table]{xcolor}
\usepackage{multirow}
\usepackage{multicol}
\usepackage{hhline}
\usepackage{wasysym} 

\def\withnotes{0} 

\definecolor{PositiveQuadrantColor}{rgb}{.13, .64, .53}
\definecolor{NegativeQuadrantColor}{rgb}{.25, .28, .53}

\definecolor{Maroon}{rgb}{0.62, 0.0, 0.09}
\definecolor{Emerald}{rgb}{.07, .74, .62}

\newcommand{\jfcolor}[1]{{\color{Emerald}#1}} 
\newcommand{\fkcolor}[1]{{\color{purple}#1}} 
\newcommand{\ascolor}[1]{{\color{violet}#1}} 

\ifnum\withnotes=1
  
  \newcommand{\as}[1]{\ascolor{\small\textbf{Aécio: }\sf #1}}
  \newcommand{\fk}[1]{\fkcolor{\small\textbf{Flip: }\sf #1}}
  \newcommand{\jf}[1]{\jfcolor{\small\textbf{Juliana: }\sf #1}}
\else
  \newcommand{\as}[1]{}
  \newcommand{\fk}[1]{}
  \newcommand{\jf}[1]{}
\fi

\definecolor{HighlightColor}{rgb}{0.05,0.05,0.70}

\definecolor{GreenLabel}{rgb}{.33, .64, .29}
\definecolor{YellowLabel}{rgb}{.93, .79, .23}
\definecolor{BlueLabel}{rgb}{.30, .47, .66}
\definecolor{OrangeLabel}{rgb}{.96, .52, .09}
\definecolor{RedLabel}{rgb}{.89, .34, .34}
\definecolor{TurkishBlueLabel}{rgb}{.45, .72, .70}

\definecolor{LightBlueLabel}{rgb}{.62, .79, .91}
\definecolor{LightOrangeLabel}{rgb}{1., .75, .47}
\definecolor{LightRedLabel}{rgb}{1., .62, .6}

\newcommand{\circleorangelight}{{\color{LightOrangeLabel} $\newmoon$}}
\newcommand{\circleredlight}{{\color{LightRedLabel} $\newmoon$}}
\newcommand{\circlebluelight}{{\color{LightBlueLabel} $\newmoon$}}

\newcommand{\circleblue}{{\color{BlueLabel} $\newmoon$}}
\newcommand{\circleorange}{{\color{OrangeLabel} $\newmoon$}}
\newcommand{\circlered}{{\color{RedLabel} $\newmoon$}}

\newcommand{\ignore}[1]{\leavevmode\unskip} 
\newcommand{\hide}[1]{}

\newcommand{\myparagraph}[1]{\vspace{0.25em}\noindent \textbf{#1.}}
\newcommand{\myparagraphem}[1]{\vspace{0.25em}\noindent \emph{#1.}}
\renewcommand{\paragraph}[1]{\vspace{0.1em}\noindent \textit{#1.}}

\newcommand{\zipcode}{ZIP Code\xspace}
\newcommand{\zipcodes}{ZIP Codes\xspace}

\newcommand{\CSK}{CSK\xspace}
\newcommand{\CSKext}{LV2SK\xspace}
\newcommand{\TUPSK}{TUPSK\xspace}
\newcommand{\PRISK}{PRISK\xspace}
\newcommand{\INDSK}{INDSK\xspace}

\newcommand{\multinom}{Mult(m, \langle p_1, p_2 \rangle)}

\newcommand{\trinomial}{\texttt{Trinomial}\xspace}
\newcommand{\cdu}{\texttt{CDUnif}\xspace}

\newcommand{\agg}{\texttt{AGG}\xspace}
\newcommand{\kmv}{\mbox{KMV}}
\newcommand{\SEQUNIQ}{\texttt{KeyInd}\xspace}
\newcommand{\SAMEX}{\texttt{KeyDep}\xspace}

\newtheorem*{defn}{Definition}

\DeclareMathOperator{\E}{\mathbb{E}}

\newcommand{\sk}[1]{\mathcal{S}_{ #1}}
\newcommand{\tb}[1]{\mathcal{T}_{ #1}}

\def\ojoin{\setbox0=\hbox{$\bowtie$}%
  \rule[-.02ex]{.25em}{.4pt}\llap{\rule[\ht0]{.25em}{.4pt}}}
\def\leftouterjoin{\mathbin{\ojoin\mkern-5.8mu\bowtie}}

\newtheorem{example}{Example}

\begin{document}

\title{Efficiently Estimating Mutual Information\\Between Attributes Across Tables}

\iftrue

\author{
  \IEEEauthorblockN{Aécio Santos}
  \IEEEauthorblockA{
    \textit{New York University}\\
    aecio.santos@nyu.edu
  }
\and
  \IEEEauthorblockN{Flip Korn}
  \IEEEauthorblockA{
    \textit{Google Research}\\
    flip@google.com
  }
\and
  \IEEEauthorblockN{Juliana Freire}
  \IEEEauthorblockA{
    \textit{New York University}\\
    juliana.freire@nyu.edu
  }
}

\fi

\maketitle

\begin{abstract}
    Relational data augmentation is a powerful technique for enhancing data analytics and improving machine learning models by incorporating columns from external datasets. However, it is challenging to efficiently discover relevant external tables to join with a given input table. Existing approaches rely on data discovery systems to identify ``joinable'' tables from external sources, typically based on overlap or containment. However, the sheer number of tables obtained from these systems results in irrelevant joins that need to be performed; this can be computationally expensive or even infeasible in practice. 
We address this limitation by proposing the use of efficient mutual information (MI) estimation for finding relevant joinable tables. We introduce a new sketching method that enables efficient evaluation of relationship discovery queries by estimating MI without materializing the joins and returning a smaller set of tables that are more likely to be relevant. We also demonstrate the effectiveness of our approach at approximating MI in extensive experiments using synthetic and real-world datasets.
\end{abstract}


\vspace{-0.1cm}
\section{Introduction}

Our increasing ability to collect and store data has led to an explosion of data repositories, both for open~\cite{nycopendata,chicagoopendata,usopendata,noy@www2019} and enterprise data~\cite{google-data-catalog,lyft-amundsen,linkedin-datahub}. 
This abundance creates opportunities to enhance data analytics and machine learning models: 
by incorporating columns from various external datasets, we can explain confounding bias~\cite{youngmann2023nexus}, test hypotheses and explain salient features in data~\cite{chirigati@sigmod2016,bessa@acmtds2020,bessa@sigmod2021}, as well as improve predictive models~\cite{chepurko2020arda,auctus@vldb2021}.
Consider the following example.

\begin{figure*}[t]
    \centering
    \includegraphics[width=1.0\linewidth]{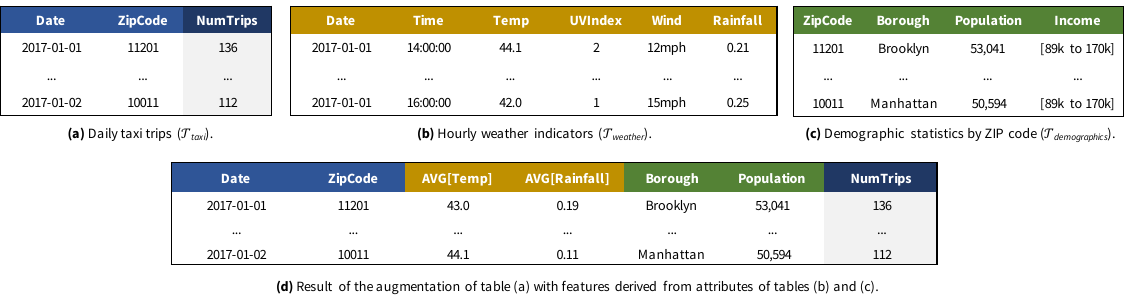}
    \vspace{-0.5cm}
    \caption{Example of relational data augmentation for the problem of taxi demand prediction. Adding new features, such as \texttt{AVG[Temp]} and \texttt{AVG[Rainfall]}, derived from external tables helps predict, or explain the variance of, the \texttt{NumTrips} attribute. The augmented table (d) is derived by joining $\mathcal{T}_{taxi}$ and $\mathcal{T}_{weather}$ on \texttt{Date}, and with $\mathcal{T}_{demographics}$ on \texttt{ZipCode}.}
        \vspace{-.5cm}
    \label{fig:taxi-demand-example}
\end{figure*}

\begin{example}[Understanding Taxi Demand]
A data scientist wants to \emph{improve a regression model for predicting taxi demand} that was constructed using historical data containing pick-up times and \zipcodes where the pick-up occurred (see table $\mathcal{T}_{taxi}$ in Figure~\ref{fig:taxi-demand-example}(a)). Here demand is measured by the total number of taxi rides ({\tt NumTrips}) originating from the same spatio-temporal region (\zipcode and Time).
Since weather is known to impact taxi demand, the data scientist obtains a new table
$\mathcal{T}_{weather}$ (Figure~\ref{fig:taxi-demand-example}(b)) that contains information about temperature and precipitation. 
By augmenting the taxi trips table $\mathcal{T}_{taxi}$ (through a join on \texttt{Date})
with hourly average temperature and hourly rainfall as additional features,
the mean absolute error of a random forest regressor improves significantly. 
In an effort to identify additional factors that may help \emph{explain the demand variability in different neighborhoods}, the data scientist joins $\mathcal{T}_{taxi}$ with $\mathcal{T}_{demographics}$ (on \texttt{ZipCode}) containing demographics statistics (Figure~\ref{fig:taxi-demand-example}(c)). The association between demand and population for the \zipcodes of pick-up locations suggests a strong dependency.
\qed
\end{example}

\vspace{-.3cm}
\myparagraph{Relational Data Augmentation} 
The manual process users must go through to discover external tables is time-consuming.
In an attempt to automate this process,  approaches have been proposed for \textit{relational data augmentation}~\cite{
zhu@vldb2016-lsh-ensemble, chepurko2020arda, fernandez@icde2019, fernandez@icde2018, zhu@sigmod2019, nargesian@vldb2018, dong@sigir2022table, nobari@icde2022efficiently}.
Most of the existing work focuses on efficient support for data integration queries, including algorithms to discover ``joinable'' tables~\cite{zhu@vldb2016-lsh-ensemble, fernandez@icde2019, fernandez@icde2018, zhu@sigmod2019, yang@icde2019, nobari@icde2022efficiently,esmailoghli2022mate},
typically based on containment or overlap.
These have an important limitation in that they can return too many irrelevant tables. Consider our example: if we search the NYC Open Data repository~\cite{nycopendata} to discover additional features that can help improve taxi demand prediction, we will find a large number of datasets that overlap in time with the taxi data, i.e., that are joinable with $\mathcal{T}_{taxi}$ on \texttt{Date}. Testing each of these by performing the join, re-training, and testing the model is wasteful and can be too expensive.

Automatic relational data augmentation systems~\cite{chepurko2020arda,ionescu@icdew2022, liu2022feature, metam} attempt to address this problem by applying feature selection techniques~\cite{vergara2014review}: given a list of joinable tables, they materialize the joins and automatically select attributes that increase the accuracy of predictive model given as input.
Because they rely on data discovery systems~\cite{fernandez@icde2018, auctus@vldb2021} to return the joinable tables, they can end up with both a large number of joins to perform and, consequently, too many candidate features to consider. Given the cost of evaluating the joins and the computational complexity of feature selection methods, it is expensive to identify the useful features.

To reduce the number of irrelevant candidate features, recent methods were proposed to discover tables that are not only joinable with a query table but also have attributes strongly correlated with an input target variable~\cite{santos@sigmod2021, esmailoghli2021cocoa, santos@icde2022}. 
Instead of joining all table pairs, these methods attain scalability by relying on inverted indexes and sketching algorithms.
While indexes allow efficient identification of tables that have join attributes with overlapping values, the sketches reduce the cost of joins and correlation computation by enabling efficient estimation of correlations between the input target variable and the top-$k$ discovered candidate attributes. 
In essence, these methods allow performing feature pruning without materializing joins, and return tables (along with their corresponding features) that are more likely to enhance model performance or provide explanations for a target variable of interest.

Unfortunately, these sketches have limitations. Notably, they do not properly handle repeated values on join keys (that are common in real data), which may cause estimation issues when performing left joins (especially on skewed distributions).
Furthermore, correlation measures may fail to identify non-monotonic relationships and they are only applicable to numerical attributes
For example, while Pearson's correlation 
may be used to identify the relationship between the numerical attributes \texttt{NumTrips} and \texttt{Rainfall},
it cannot be directly applied to \texttt{Borough}, which is a categorical attribute.
It also can fail to identify the non-monotonic relationship between \texttt{NumTrips} and \texttt{Population},
assuming taxis have fewer pick-ups both in neighborhoods with small populations (due to fewer customers) and large populations (due to heavy traffic). 

\myparagraph{Beyond Correlation-Based Data Discovery}
Ideally, we would like to use a more general measure of statistical dependence, such as \textit{Mutual Information (MI)}, which is invariant under homomorphism. 
Because MI is defined over probability distributions (for scalars and vectors), it is applicable to multiple data types and dimensionalities.
Due to its generality, MI has found applications in numerous problems including the analysis of gene expression~\cite{daub2004estimating}, functional dependency discovery~\cite{mandros2017discovering, mandros2019discovering, mandros2020discovering, pennerath2020discovering}, explanatory data analysis~\cite{youngmann2022explaining}, causality detection~\cite{causality-detection}.
In machine learning, MI is also used in many feature selection methods~\cite{chandrasekar@2014,feature-selection-acmsur2017, vergara2014review}.
Moreover, strong theoretical connections between MI and model generalization error have been found showing that regression and classification errors are minimized when features having the largest conditional MI with the target are selected~\cite{beraha2019feature, peng2017feature, brown2012conditional}.

\myparagraph{Challenges of MI Estimators}
Estimating MI from finite data samples is far from trivial. Commonly used MI estimators based on empirical entropy do a poor job of modeling the underlying distribution due not only to small sample sizes but also to inherent estimator bias~\cite{roulston1999estimating} (see Section~\ref{sec:background}). 

Moreover, while MI is well-defined for different data types, existing estimators typically either handle only discrete-categorical or continuous-numerical data.
A common approach to handle continuous data is to discretize it using binning techniques~\cite{hacine2018binning} and then use an estimator for discrete data.
Unfortunately, binning makes strong assumptions about the data distribution, may lead to information loss,
and has bias that increases with the number of distinct values for the maximum likelihood estimator (MLE) on discretized data~\cite{paninski2003estimation}.

Instead of binning numerical data, some implementations \cite{scikit-learn} employ different estimators depending on the attribute data type.
Specifically, they use the MLE estimator
(i.e., maximum likelihood estimator plug-in of the empirical distribution~\cite{7060676}) 
for categorical/discrete data and the KSG estimator for numeric~\cite{kraskov2004estimating} or mixed numeric-categorical~\cite{ross2014mutual} values.
While this approach avoids the pitfalls of data transformation techniques, it is not clear if using this combination of estimators with different properties (e.g., bias) is more effective.

\myparagraph{MI Estimation over Joins}
In the particular setting of relational data augmentation, these issues are compounded by additional problems created by many-to-one left joins and the need for efficiency.
In this scenario, the goal is to augment a given base table with additional features that will be used to train a machine learning model to better predict or explain a target variable. Therefore, we need to keep the number of rows in the original table intact via a left-outer join.
However, this creates some issues.

First, joining tables on non-unique join-key attributes leads to the creation of feature attributes containing repeated values that follow the distribution of the join key.
For instance, while the original \texttt{Population} attribute in Figure~\ref{fig:taxi-demand-example}(c) may have unique values, the derived \texttt{Population} attribute in the augmented table shown in Figure~\ref{fig:taxi-demand-example}(d) will have additional repeated entries according the distribution of the \texttt{ZipCode} join key.
In the particular case where the external attribute has a continuous distribution, the derived feature attribute will be a mixture of continuous distributions (with repeated values) that need to be handled properly by specialized MI estimators~\cite{gao2017estimating}.

Second, to avoid the cost of fully materializing joins, systems are limited to estimating MI using a small number of samples of the join obtained using sketches~\cite{santos@sigmod2021, huang2019joins}.
State-of-the-art methods typically use coordinated sampling (based on minwise hashing) to increase the number of samples that contribute to the join~\cite{cohen2023sampling}. 
However, given that coordination is typically achieved by hashing values of the join-key attributes, these algorithms may introduce sampling bias and dependence on the join-key that leads to violating the i.i.d. assumptions of estimators.
When this happens, the bias of MI estimators is further increased (as shown in Sections~\ref{sec:sketches-tuple-sampling} and~\ref{sec:experiments}).

Finally, existing join-sampling algorithms~\cite{vengerov@vldb2015, chen@sigmod17, huang2019joins} are not directly applicable as they
use estimators that are specialized for the specific function they aim to estimate, typically \texttt{COUNT} or \texttt{SUM}.
Moreover, they have multiple sampling rate parameters that are difficult to set in practice, and do not provide a bound on sketch size.

\myparagraph{Contributions}
We define the problem of MI estimation over joins for data augmentation, and identify challenges that arise when estimating it over left joins with non-unique keys.
We propose a new sketch that has a single parameter, the maximum sketch size, and addresses MI estimation challenges while avoiding the full join computation, thus providing efficient support of MI-based data~discovery. Unlike previous approaches, our sketching method: does not assume that join keys are unique, and guarantees a fixed-size sketch while keeping an unbiased uniform sample of the join and thus can be used with any existing sample-based MI estimator.

We assess the effectiveness of our sketching method and different MI estimators through an extensive experimental evaluation.
To do so, we design a synthetic benchmark that allows us to compare the estimated and the true MI obtained analytically from the data distributions used to generate the data.
This benchmark allows us to observe the impact of dependence between the join-key attributes and the feature attributes on the MI estimates computed using the sketches.
Furthermore, it allows us to identify 
differences in the behavior of various combinations of sketches and MI estimators, when and why they fail.
We also evaluate our sketches using real-world data from open-data repositories.
Our results confirm that the proposed sketch enables efficient approximation of the MI computed on the full join, and uncovered useful observations that help guide the implementation and deployment of MI-based sketches for data augmentation.

\vspace{-.15cm}
\section{Background}
\label{sec:background}

Before presenting our approach, we first present the terminology used in the paper and some information-theoretic measures such as entropy and mutual information.

\myparagraph{Data Types}
As a simplification, we shall use the terms \textbf{discrete} and \textbf{continuous} to distinguish types of value distributions,
with the former reserved for what is referred to in the literature as (often unordered) categorical, and the latter as (ordered, often floating-point) numerical~\cite{shah21featureinference}.
We are aware that real data is more complicated and may include integral categories (e.g., UPC code) and floating-point values that represent discrete categories (e.g., Dewey Decimal).
We assume such cases are represented as strings in Section~\ref{sec:experiments}.

It is also important to differentiate a single \textbf{mixture} attribute, that contains a mixture of continuous distributions
(e.g., the variable \texttt{AVG[Temp]} from the weather table in Figure~\ref{fig:taxi-demand-example}(d) has repeated values for each zip code on the same date),
from a pair of attributes where each contains a different data type.

\myparagraph{Entropy}
\emph{Entropy} quantifies the amount of ``information'' or ``uncertainty'' in the possible outcomes of a random variable.
Let $X$ be a discrete random variable that assumes values from $dom(X) = \{1, ..., u_X\}$ and has \textit{probability mass function} $p(X)$. The entropy $H(X)$ of the random variable $X$ is: 
\vspace{-.2cm}
\begin{equation} \label{eq:entropy}
\vspace{-.25em}
H(X) = \E[- \log p(X)] = - \sum_{i=1}^{u_X} p(i) \log p(i)
\vspace{-.25em}
\end{equation}
Analogously, when $X$ is a continuous random variable whose support is defined over the set $\mathcal{X}$ and has probability density function $f(X)$, the \textit{differential entropy} is defined~as:
\vspace{-.25em}
\begin{equation} \label{eq:differential-entropy} 
H(X) = \E[- \log f(X)] = - \int_{\mathcal{X}} f(x) \log f(x) \,dx
\vspace{-.25em}
\end{equation}

These measures can be generalized to multiple variables. Let $X$ and $Y$ be random variables, whose support values are the ranges $[1, u_X]$ and $[1, u_Y]$, respectively, with  joint probability mass function $p(X, Y)$. We define the \textit{joint entropy}~as: \vspace{-.2cm}
\begin{align}
H(X, Y) = - \sum_{i=1}^{u_X} \sum_{j=1}^{u_Y} p(i,j) \log p(i,j)
\end{align}
Analogously, if $X$ and $Y$ are continuous with support on $\mathcal{X}$ and~$\mathcal{Y}$, respectively, and joint probability density function $f(x, y)$, the \textit{joint differential entropy} is defined as:
\begin{align}
H(X, Y) = - \int_\mathcal{X,Y} f(x,y) \log f(x,y) \,dx\,dy
\end{align}

\myparagraph{Mutual Information (MI)}
\noindent The MI between two variables $X$ and $Y$ quantifies the amount of information obtained about one variable by observing the other. It is defined as:
\vspace{-.1cm}
\begin{align}
I(X, Y) 
&\triangleq H(X) + H(Y) - H(X, Y) \label{eq:mutual-info}
\vspace{-.1cm}
\end{align}
Intuitively, MI represents the amount of information learned about the target variable $Y$ after observing the feature $X$.
It is widely used in applications including feature selection~\cite{beraha2019feature}, data augmentation~\cite{liu2022feature}, and causality analysis~\cite{4738640}.
In decision tree learning, it is known as \textit{information gain}.

Note that $I(X, Y) \geq 0$, with $I(X, Y) = 0$ only when $X$ and $Y$ are independent variables.
What makes MI particularly attractive is its robustness due to invariance under reparameterizations:
any bijection on discrete values and any homomorphism on continuous values, including affine transformations, has the same MI \cite{kraskov2004estimating}.

\hide{
Moreover, MI is biased toward high-cardinality attributes. For example, using MI as a criterion to select the next feature when building a decision tree biases the algorithm towards selecting high-cardinality attributes, resulting in a large number of branches and lower accuracy \cite{quinlan1986induction}. To reduce this bias, \citet{quinlan1986induction} proposed the \textbf{information gain ratio} (IGR) as a criterion to select attributes to split trees:
$IGR(X, Y) = I(X, Y)/H(X)$.
By normalizing the MI by the entropy of the feature $X$, the algorithm reduces the bias towards high-cardinality attributes and improves
accuracy \cite{quinlan1986induction}.
%
Other examples of normalized mutual information (NMI)
include $\frac{I(X, Y)}{\min\{H(X), H(Y)\}}$, $\frac{I(X, Y)}{\max\{H(X), H(Y)\}}$ and $\frac{I(X, Y)}{\sqrt{H(X) H(Y)}}$ (we refer the reader to \cite{vinh2010information} for a detailed comparison of these and other measures).
}

\myparagraph{Estimating Entropy and Mutual Information}
%
The measures above are defined over probability distributions.
However, the distribution is usually unknown in real-life applications.
Therefore, applications often use an estimator based on a finite number of observations to approximate the distribution.

The classical \textit{maximum likelihood estimator (MLE)} of entropy is obtained by estimating the probability mass function using frequencies as follows.
Given attribute $X$, let $N$ be the number of observations in $X$
and $N_i$ be the frequency of each element $i$ in $X$ (hence, $N = \sum_{i=1}^{u_X} N_i$).
The empirical entropy is estimated as:
\vspace{-.3cm}
$$
\hat{H}_{MLE}(X) = \sum_{i=1}^{u_X} \frac{N_i}{N} \log \frac{N_i}{N}.
\vspace{-.2cm}
$$
The MLE estimator is known to be systematically biased downward from the true entropy, and
the bias is influenced by the number of samples $N$ and distinct values $m_X$, and the distribution of~$X$~\cite{roulston1999estimating}.

The MLE estimator is only applicable to discrete variables.
To estimate entropy over values from a continuous domain $X$, one can estimate $H(X)$ from the average distance to the $k$-nearest neighbor ($k$-NN), averaged over all $x_i \in X$.
It is well known that $H(X) \approx \frac{1}{N-1} \sum_{i=1}^{N-1} \log(x_{i+1}-x_i) + \psi(1) - \psi(N)$, where $\psi$ is the digamma function~\cite{kraskov2004estimating}.

The \textit{mutual information} (MI) can be obtained by estimating $H(X)$, $H(Y)$, and $H(X,Y)$ separately and calculating Equation~\ref{eq:mutual-info}. When both $X$ and $Y$ are discrete, estimates can be obtained using the MLE estimator. When both of them are continuous, MI can be obtained using the KSG estimator \cite{kraskov2004estimating} which computes $I(X,Y)$ in a slightly different way to avoid compounding errors of the terms.
Alternatively, if both components are mixtures of discrete-continuous distributions, the MixedKSG estimator~\cite{gao2017estimating} can be used. MixedKSG proceeds similarly to KSG but recovers the plug-in estimator in discrete regions of the distribution (if they exist).
Finally, when components have different types of distributions (i.e., discrete-continuous or continuous-discrete cases), another variation of the KSG estimator can be used \cite{ross2014mutual}: first, the $k$-NN distances are computed for each discrete value using only the continuous variable, then the cardinality among all continuous values for those distances is calculated.

These MI estimators have different biases. The MLE estimator (for the discrete-discrete case) has a bias proportional to the number of distinct values and sample size~\cite{roulston1999estimating}:
\vspace{-.1cm}
\begin{align} \label{eq:mi-mle-bias}
I(X, Y) - \E[\hat{I}_{MLE}(X, I)] \approx
\frac{m_X + m_Y - m_{XY} - 1}{2N}
\end{align}
The KSG estimators, on the other hand, have a bias that stems from uniformity assumptions on the density and depends on neighbor distances~\cite{kraskov2004estimating}. 
In Section~\ref{sec:experiments}, we provide an experimental comparison of these estimators on multiple datasets.

\vspace{-.15cm}
\section{MI Estimation For Data Augmentation}
\label{sec:relational-data-augmentation}

We are interested in estimating mutual information in the relational data augmentation setting: augment a given base table with new attributes from an external table through a join operation. Since these new attributes may be used
as additional features to train a machine learning model to predict a target variable, we need to  keep the number of rows in the original base table intact by 
performing a left-outer join.

\vspace{-.15cm}
\subsection{Problem Statement} \label{sec:problem-statement}
Let $\tb{train}$ denote the base table containing (1)~a target variable $Y$ that we want to predict or explain, and (2) an attribute $K_Y$ (or set of attributes) that can be used as a join key in a relational equi-join operation.
Let $\tb{aug}$ be an external table that contains (1) an attribute $X$ (or a set of attributes) that can be used as a feature; and (2) an attribute $K_X$ that can be used to join  $\tb{aug}$ with the base table $\tb{train}$ on the attribute $K_Y$.
This is formally defined below. 
%
For exposition, we assume the case where $K_X$, $K_Y$ and $X$ are all single attributes,
and we discard any rows with \texttt{NULL} values resulting from $\tb{aug}$ not containing some key $k$ in $K_X$ that is present in $K_Y$.
\footnote{
While we limit joins to having high containment,
our method does not prevent using existing strategies for handling \texttt{NULL}s in MI estimation~\cite{DOQUIRE20123,hutter2005distribution}.
Evaluating these approaches is beyond the scope of this paper.
}

\vspace{-.1cm}
\begin{defn}{(\bf MI Estimation over Joins)}
Given two tables $\tb{train}$ and $\tb{aug}$,
the goal is to estimate the mutual information between the attributes $X$ and $Y$
from the table constructed through a left-outer-join of the two tables, $\tb{train} \leftouterjoin \tb{aug}$, on keys $K_X$ and $K_Y$, without having to compute the join.
\end{defn}
\vspace{-.1cm}

Consider the example in Figure~\ref{fig:taxi-demand-example}. Here, $\tb{train}=\tb{taxi}$ is a table containing the number of daily taxi trips in New York City (NYC).
The attribute $Y$ represents the number of taxi trips \texttt{NumTrips} that happened in a particular \zipcode with $K_X=K_Y=\texttt{ZipCode}$. We are interested in enriching $\tb{train}$ with new features that may help predict the taxi demand (number of trips) on a given day. $\tb{aug}=\tb{demographics}$ represents another table discovered in a different source.
To determine if $\tb{aug}$ may be useful for our predictive task, we want to estimate the mutual information between the columns $X$ and $Y$ (e.g., \texttt{Borough} and \texttt{NumTrips}), obtained after the join between $\tb{train}$ and~$\tb{aug}$, without computing the full join.


\hide{
\begin{table}[t]
\begin{tabular}{cc}
\toprule
\textbf{Symbol} & \textbf{Meaning} \\ \midrule
$X, Y$          & feature and target attributes \\
$K_X (K_Y)$     & key attribute in the same table as $X$ (resp. $Y$)         \\
$dom(X)$        & set of values in attribute $X$ \\
$N$             & number of observations                 \\
$N_i$           & number of observations with value $i$          \\
$m_X$           & number of distinct values in $X$ \\
$\tb{train}$    & training table   \\
$\tb{aug}$      & augmentation table \\
$\tb{cand}$      & candidate table \\
$n$             & sketch size parameter \\ 
\bottomrule
\end{tabular}
\caption{Notation summary.}
\vspace{-.5cm}
\label{table:notation-summary}
\end{table}
}

\subsection{Joining Arbitrary Tables}
\label{sec:mi-estimation-general-case}

\myparagraph{Many-to-Many Joins}
Our problem definition assumes that there is a many-to-one relationship between $\tb{train}$ and~$\tb{aug}$,
that is, each tuple from $\tb{train}$ joins with at most one tuple from $\tb{aug}$.
This is required by applications such as model improvement, where the number of rows in the original training set must remain intact to avoid introducing bias.
Furthermore, if new rows are added during the join, it is not clear which labels should be assigned to these rows.
However, while searching for augmentations, we can find candidate tables $\tb{cand}$ that have a many-to-many relationship with $\tb{train}$. For example, when joining taxi trips $\tb{taxi}$ and weather $\tb{weather}$ on \texttt{Date}, since temperature \texttt{Temp} values are recorded at hourly intervals, there are multiple temperature readings associated with each \texttt{Date}. 

In such cases, since the augmentation will lead to duplicate key values, we transform the candidates to ensure that the augmented table will have the same number of rows as $\tb{train}$.
To do so, we define a \textit{featurization} function \agg that derives the augmentation table $\tb{aug}$ from a candidate table $\tb{cand}$.
Given a candidate table $\tb{cand}$ with key column $K_Z$ and value column $Z$, and a featurization function \agg,
the following join-aggregation query maps $\tb{cand}[K_Z, Z] \rightarrow \tb{aug}[K_X, X]$, generating an intermediate aggregate table $\mathcal{T}_{aug}$ which is then combined with $\tb{train}$. This can be expressed using the following SQL query:

\begin{adjustwidth}{1.5em}{1em} 
\small
\vspace{.3em}
\texttt{\noindent
\textbf{SELECT} $\tb{train}$[$K_Y]$, $\tb{train}[Y]$, $\tb{aug}[X]$ \\
\textbf{FROM} $\tb{train}$ \\
\textbf{LEFT JOIN} ( \\
\hspace*{.5em} \textbf{SELECT} $K_Z$ \textbf{AS} $K_X$, AGG(Z) \textbf{AS} X from $\tb{cand}$ \\
\hspace*{.5em} \textbf{GROUP BY} $K_Z$ \\
) \textbf{AS} $\tb{aug}$ \\
\textbf{ON} $\tb{train}[K_Y]$ = $\tb{aug}[K_X]$;
}
\vspace{.3em}
\end{adjustwidth} 
Note that while the aggregation of $\tb{cand}$ can be performed separately as a preprocessing step, the materialization of $\tb{aug}$ is not required for MI estimation.
As we will discuss in Section~\ref{sec:mi-estimation}, sketches can be constructed directly from $\tb{cand}$, which avoids the cost of aggregating join keys that are not needed for estimating the MI.

Figure~\ref{fig:taxi-demand-example} illustrates an example where the attribute $Z=\texttt{Temp}$ in $\tb{cand}=\tb{weather}$ is transformed (i.e., the values associated with a given date averaged) and joined on $\texttt{Date}$ with $\tb{train}=\tb{taxi}$, as shown in the resulting table in Figure~\ref{fig:taxi-demand-example}(d).
Note that the numbers of rows in the tables $\tb{train}$ and $\tb{cand}$ need not be equal and their join attributes may have repeated values that need to be mapped to a feature value. We give a more concrete example below.

\begin{example} \label{ex:data-aug-join-example}
Let $K_Y$ and $K_Z$ be the keys for the tables $\tb{train}$ and $\tb{cand}$, respectively. Let $\tb{train}[K_Y] = [a, a, b, c]$ 
and $\tb{cand}[K_Z] = [a, b, b, b, c, c, c]$ and $\tb{cand}[Z] = [1, 2, 2, 5, 0, 3, 3]$, respectively.
The first step is to group values based on the key values, i.e., $\{a \rightarrow [1], b \rightarrow [2,2,5], c \rightarrow [0,3,3]\}$. 
Next, the aggregate function \texttt{AVG} generates an intermediate aggregate table $T_{aug}$ with the mappings $[a \rightarrow 1, b \rightarrow 3, c \rightarrow 2]$. 
Joining $T_{aug}$ with the training table $T_{train}$ generates the column $X=[1, 1, 3, 2]$.
Similarly, if we applied \texttt{MODE} (to return the most frequent value), the output would be $X = [1, 1, 2, 3]$, and \texttt{COUNT} would generate $X = [1, 1, 3, 3]$.
\end{example}
\vspace{-.2cm}

From the example above, we can draw a few observations about featurization and its implications for MI estimation. 

\myparagraphem{Data Distribution} 
The distribution of the feature $\tb{train}[X]$ depends on the function \agg and the join key $\tb{train}[K_Y]$.
For instance, distribution parameters such as the mean of $X$  will likely be different for functions such as \texttt{AVG} and \texttt{MAX}. Note also that repeated values in $K_Y$ lead to repeated values in~$X$, e.g., the value $1$ is repeated twice in $X$ because $a$ repeats twice in $K_Y$. Finally,  note that $\tb{train}[X]$ may be even independent of $\tb{cand}[Z]$ when using a function such as \texttt{COUNT}, in which case it only depends on the key frequency distribution of $K_Y$ (assuming no \texttt{NULL} values exist in $Y$).

\myparagraphem{Choice of Aggregation Function} 
There are many choices for aggregate functions and some may be more appropriate for specific data types, e.g., \texttt{AVG} for ordered-continuous data versus \texttt{MODE} for unordered-discrete data. Furthermore, the data type of the function output depends on the aggregation function and the input data type (e.g., while \texttt{COUNT} always outputs a discrete number regardless of input type, \texttt{MODE} outputs the same type as the input data.
Note also that the aggregation function is not limited to outputting scalar values.
It could also return multidimensional vectors (such as embeddings of text values),
in which case, any MI estimator that supports multidimensional variables such as KSG~\cite{kraskov2004estimating} and its derivatives could be used (see Section~\ref{sec:background}).
In practice, to support general datasets and information needs, it may be necessary to create multiple augmentation columns using different aggregation functions and then examine the results.

\section{MI Estimation using Sketches}
\label{sec:mi-estimation}
\vspace{-.2cm}
The task of estimating quantities over join results without materializing the join is a long-standing problem in the literature~\cite{acharya@sigmod1999joinsynopses, huang2019joins}.
%
%
%
%
One way to do so is to use sampling. A na\"ive approach to obtain samples of an equi-join is to join rows sampled \textit{independently} from the two tables via Bernoulli sampling. Unfortunately, this results in a quadratically smaller join size~\cite{acharya@sigmod1999joinsynopses} which results in poor accuracy.

To address this issue, we use \textit{coordinated sampling} \cite{huang2019joins, vengerov@vldb2015, chen@sigmod17, BessaDFMMSZ:2023, beyer@sigmod2007, santos@sigmod2021, santos@icde2022, DBLP:reference/algo/Cohen16a, cohen2023sampling}.
In this approach, a shared-seed random hashing function is applied to the join keys, and a small set of tuples with the minimum hash values is selected to be included in the sketch~\cite{huang2019joins, BessaDFMMSZ:2023}.  
This implies that if a row with join key $k$ is chosen from table~$\tb{train}$, then a corresponding row with the same key $k$ in~$\tb{aug}$ is more likely to be sampled.
Hence, we essentially forgo sample independence for an increased join size. However, this may lead to samples that are not identically distributed, which increases estimator bias. 
While there exist weighting schemes such as Horvitz-Thompson for correcting this bias~\cite{daliri2023sampling, estan@icde2006endbiased}, these estimators are tightly coupled with the measures being estimated and, thus, are not directly applicable to estimating MI (see Section~\ref{sec:background}).
We, therefore, focus on sampling schemes that allow us to use existing MI estimators.

Another challenge is how to deal with repeated values in the join key. Coordinated sampling, which chooses samples based on hash values of join keys, typically assumes that the keys are unique or, if not, can be aggregated~\cite{santos@sigmod2021, BessaDFMMSZ:2023, vengerov@vldb2015}. As discussed in Section~\ref{sec:relational-data-augmentation}, this does not hold in the relational data augmentation setting since the number of rows must be kept intact. To address this issue, some join sampling algorithms suggest an all-or-nothing approach that includes all entries associated with a selected join key~\cite{vengerov@vldb2015}. However, this is known to increase the variance of estimators and results in unbounded size~\cite{chen@sigmod17}. Recent approaches introduce multiple sampling-rate parameters to select a fraction of the repeated entries~\cite{chen@sigmod17, huang2019joins}. While this improves variance, these parameters are hard to set in practice and the resulting size is sensitive to table size and the join key distribution.

We propose new sampling-based sketches for estimating MI over joins that address these challenges.
We start by proposing an extension of existing two-level sampling schemes~\cite{huang2019joins, chen@sigmod17} that takes only a single parameter as input and provides a hard bound on sketch size (Section~\ref{sec:csk-two-level}). We refer to this baseline approach as \CSKext.
We provide an analysis that shows that \CSKext leads to non-uniform sampling (hence, not identically distributed) that can increase the bias of MI estimators; this is confirmed experimentally in Section~\ref{sec:experiments-join-key-effect}. 
To address this issue, we propose \TUPSK (Section~\ref{sec:sketches-tuple-sampling}), a novel tuple-based sampling scheme that leads to uniform sampling probabilities, which better matches assumptions made by the MI estimators. 
As shown in Section~\ref{sec:experiments-join-key-effect}, this reduces bias and leads to higher accuracy of MI approximation.

\vspace{-.1cm}
\myparagraph{Approach Overview}
We assume access to hash function $h_u$ that maps input uniformly to the unit range $[0,1]$.
We also assume that inputs to $h_u$
are integers. Otherwise, 
we transform the input to integers using a collision-free hash function $h$ that maps objects to integers before feeding them to $h_u$, i.e., $h_u(h(x))$ where $x$ is the input data. 
In practice, it suffices to use pseudorandom functions. To implement $h$, we used the well-known 32-bit MurmurHash3 function. For $h_u$, we used Fibonacci hashing \cite{knuth1997art}. 

Our approach works as follows.
Given tables $\tb{X}$ and $\tb{Y}$ with schemas $[K_X, X]$ and $[K_Y, Y]$ respectively, we first build small sketches $\sk{X}$ and $\sk{Y}$.
The sketch $\sk{X}$ is composed of~a set of tuples $\langle h(k), x_k \rangle$ where $h(k)$ is the hash of a join key value $k \in K_X$ and $x_k$ is a value from $X$; the sketch $\sk{Y}$ is built analogously to $\sk{X}$.
Sketches are typically built in an offline preprocessing stage. When it is time to estimate MI between attributes in two different tables, we merge their sketches to recover a useful sample of the join for estimating the mutual information.
Specifically, given a pair of sketches $\sk{X}$ and $\sk{Y}$, we create a sketch $\sk{join}$ by performing a join between the sketches on their hashed keys~$h(k)$, resulting in tuples $\langle h(k), x_k, y_k \rangle$. 
These tuples are a subset of the full table join $\tb{join}$.
Finally, we apply a function~$\mathcal{F}$ that uses the sample of paired values $\langle x_k, y_k \rangle$ in $\sk{join}$ to estimate the MI of $X$ and $Y$, i.e., $\hat{I} = \mathcal{F}(\sk{join})$.
The function $\mathcal{F}$ uses existing MI estimators such as the ones described in Section~\ref{sec:background}.

Our sketching algorithms only differ in the strategy they use to select samples that are included in the sketch. 
As inputs, they are given tables $\mathcal{T}_{train}$ (left table) and $\mathcal{T}_{cand}$ (right table). 
Specifically, when sketching $\mathcal{T}_{train}$, it must sample values associated with repeated key values, whereas, for $\mathcal{T}_{cand}$, it must aggregate repeated values in order to create a sketch that represents $\mathcal{T}_{aug}$. In what follows, we provide a detailed description of these methods.

\vspace{-.2cm}
\subsection{Baseline: Two-Level Sampling (\CSKext)}
\label{sec:csk-two-level}
\vspace{-.1cm}
Our first sketching method works as follows.
In the first level, it performs coordinated sampling based on (distinct) join keys to select the same set of keys from both tables, thus maximizing the expected join size between $\tb{X}$ and $\tb{Y}$.
However, given that this does not provide a bound on the sketch size, it performs a second sampling step to cap the number of samples per key, limiting the final sketch size.

\myparagraph{Building \CSKext Sketches}
We choose the set of tuples $\langle h(k), x_k \rangle$ as follows.
At the first sampling level, we select the $n$ keys that have the minimum values of $h_u(k)$. For each of these $n$ keys, we filter a subset of the tuples having key $k$ using independent Bernoulli sampling. The number of tuples to be included in the sketch is chosen in proportion to the frequency of $k$ in the original table $\tb{}$, as follows:
\begin{enumerate}[leftmargin=15pt]
    \item For $\tb{cand}$, we apply aggregate function \agg to the set of values $\{x_k\}$ associated with each key $k$ to generate a single value $\agg(\{x_k\})$.
    \item For $\tb{train}$, we keep $n_k = \max(1, \lfloor n p_k \rfloor)$ samples per key, where $p_k = N_k/N$ is the probability of the key $k$ in $K_X$.
\end{enumerate}

\noindent The sampling strategy above guarantees that
(1) the sketch contains at least one sample for each of the $n$ chosen keys, and
(2) for the chosen keys, the frequency of the key in the sketch is proportional to the frequency of the key in $\tb{train}$.
We can build the sketch described above after sorting $\tb{train}$.
Alternatively, it can be done~in~a single pass using reservoir sampling:  we only need to maintain a reservoir with $n$ samples for each of the $n$ minimum keys and the number of repeated entries associated with each of the minimum keys,
which is needed to determine the desired number samples $n_k$ for each join key~\cite{vitter1985random}.
At the end of this pass, we keep only the first $n_k$ samples of the reservoir of each key $k$ and discard the remaining entries.

\myparagraph{Sketch Size}
The size of \CSKext sketches is upper bounded by $2n$, but it is typically close to $n$;
To see this, note that the sketch size is given by: $\sum_{k_i \in \kmv(K_X)} n_i = \sum_{k_i \in \kmv(K_X)} \max\left(1, \left\lfloor \frac{n N_i}{N} \right\rfloor\right)$
where $k_i \in \kmv(K_X)$ denotes the set of $n$ minimum values in $K_X$ selected in the first-level sampling.
It is easy to show that the upper bound for its size is $2n$, and $\sum_{k_i \in \kmv(K_X)} n_i \geq n$ holds whenever the number of unique values in the join key $m_{K_X} \ge n$. 

\myparagraph{Analysis}
Let $p_i = \Pr[t_i]$ be the probability of selection of one tuple $t_i$ in the first sampling level,
and $q_i = \Pr[t_i]$ be the selection probability of $t_i$ in the second-level.
Since we assume a uniform hashing function $h_u$, in the first level any join-key value $t_i[K]$ has a uniform inclusion probability $p_i = 1/m_K$, where $m_K$ is the number of distinct values in $K$.
In the second level, where we perform Bernoulli sampling of $N_i$ tuples that were sampled in the first level, we have that $q_i = 1/\max\left(1, \left\lfloor n \frac{N_i}{N} \right\rfloor \right)$.
Given that the inclusion in the second level is independent of the first, the final probability of selecting the tuple $t_i$ is
$
\Pr[t_i] = p_i q_i = 1 / \left( m_K \cdot \max\left(1,  \left\lfloor n \frac{N_i}{N} \right\rfloor \right) \right)
$.
Here we can see that the tuple selection probability is clearly dependent on the frequency distribution of the join-key values.
Note, however, that in the special case where join keys are unique, i.e., $m_K=N$, the sampling probability becomes uniform as $\Pr[t_i] = 1/\left(N \cdot \max\left(1, \left\lfloor n \frac{1}{N} \right\rfloor\right) \right) = 1/(N\cdot1) = 1/N$.
An example of this arises when creating the sketch $\sk{aug}$, as it always aggregates the keys to a single value.

\subsection{Proposed Approach: Tuple-based Sampling~(\TUPSK)}
\label{sec:sketches-tuple-sampling}

\CSKext sketches have important limitations.
When a join key $k$ is not selected for inclusion in the sketch in the first-level sampling, none of the rows that contain $k$ will be included in the sample. 
Moreover, the sampling of a join key value does not take into account its frequency in the table.
To see why this is a problem, consider the following extreme example.

Assume that we have a table $\tb{train}[K_Y,Y]$ of size $N=100$. Let $K_Y=[a,b,c,d,e,f,f,f,...,f]$ and $Y = [0,0,0,0,0,1,2,3,...,95]$.
Given that $\Pr[Y=0]=0.05$ and $\Pr[Y=i]=0.01$ when $i \neq 0$, we have that the entropy of $Y$ is $\hat{H}(Y) = -0.05 \log(0.05) - {95} \times 0.01 \log(0.01) \approx 4.5247$.
Now consider a \CSKext sketch of size $n=5$. In the case that the keys $a,b,c,d,e$ are selected in the first-level sample, then $\sk{train}[Y] = [0,0,0,0,0]$, and thus its entropy estimate $\hat{H}(Y) = -1 \log 1= 0$.
Given that the entropy upper-bounds MI, we also have that the MI estimate between $\sk{train}[Y]$ and any feature column $X$, regardless of what its values are, must also be 0.
Additionally, note that the probability of selecting a tuple $t_i$ such that the key value is equal to $f$, $\Pr\left[t_i[K_Y]=f\right]$, is $1/(6\max(1, 4.75)) = 0.035$ (since $n\frac{N_i}{N} = 5\frac{95}{100} = 4.75)$, whereas for each of the other key values is $1/(6\max(1,0.05)) \approx 0.167$.
This example illustrates how dependence between the join key and the target attribute can lead to estimation bias.

To address these problems, we propose a coordinated sampling scheme that (1) considers individual rows as a sampling frame (i.e., each row is considered for sampling individually) and (2)  leads to identically distributed samples (i.e., each row has the same probability of being sampled).
Moreover, given that the probability of sampling each row is uniform, the expected number of sampled rows that contain a given join key $k$ is proportional to the frequency of $k$ in the original table. 
We refer to this method as \TUPSK.

\myparagraph{Building \TUPSK Sketches}
To build \TUPSK sketches, we select rows from $\tb{train}$ by hashing keys as follows.
We use the tuple $\langle k, j \rangle$ to identify the row where $k$ appears for the $j^{th}$ time in sequence, resulting in derived keys $\langle k, 1\rangle $, $\langle k, 2\rangle $, $...$, $\langle k, N_{k}\rangle$. Then, instead of selecting the rows based on the minimum hash values of $h_u(k)$, we select tuples based on $h_u(\langle k, j\rangle)$. The final sketch, however, stores only tuples containing the hashed key and its associated value: $\langle h(k), x_k \rangle $. In other words, the tuples $\langle k, j \rangle $ are only used for deciding whether or not to include a row in the final sketch $\sk{train}$.

When sketching $\tb{cand}$, we handle repeated values as in \CSKext: we apply $\agg$ to the set $\{x_k\}$, of values from $X$ appearing with key $k$, to derive a single value $x_k = \agg(\{x_k\})$.
Then we select tuples with the $n$ minimum values of $h_u(\langle k, 1\rangle)$, since the aggregation results in unique keys. Hashing on $\langle k, 1 \rangle$ provides sample coordination between the sketches $\sk{train}$ and $\sk{aug}$.

\myparagraph{Analysis} Let $p_i$ be the probability of selecting a tuple $t_i$ to be included in a \TUPSK sketch. 
It is easy to see that each $\langle k, j \rangle$ uniquely identifies a row in the table.
Since $h_u$ is a uniform hashing function, $p_i = 1/N$.
Note that, unlike \CSKext sketches, the probability is uniform regardless of the frequency distribution of the join keys.
Note also that, for the particular case of data augmentation, the tuples $\langle k, j \rangle$ also uniquely identify the rows in the final left join since the join is many-to-one and its output has the same size as the left table. Hence, the sample recovered by a sketch join is a uniform sample of the full join result.

It is worth noting that not all samples in the sketch are coordinated. This happens because the aggregation of tuples with repeated keys
limits the domain of the tuples 
to $\langle k, 1\rangle$.
In contrast, when sketching $\tb{train}$, the domain of the tuples $\langle k, j\rangle$ depends on the frequency distribution of the join key $\tb{train}[K_Y]$.
This implies that the hashes of all tuples from $\sk{train}$ having $j>1$ cannot match any tuples from $\sk{aug}$. Consequently, the sampling of such tuples is equivalent to a Bernoulli sampling.

Finally, note that unlike in \CSKext, each repeated key in $\tb{train}$ may evict other tuples from the $n$ minimum values in $\sk{train}$. While somewhat counter-intuitive, less coordination means higher sample quality as this reduces dependence on the join keys (i.e., the sample becomes closer to an independent Bernoulli sample).
Overall, our experimental results show that this scheme leads to a better trade-off between coordination and independence (Section~\ref{sec:experiments-synthetic-data}).

\myparagraph{Accuracy Guarantees} 
\TUPSK provides unbiased uniform samples of the join, but the actual estimation accuracy guarantees provided by our sketch also depend on the selected MI estimator used. 
All MI estimators used in this paper have been proven to be consistent estimators~\cite{paninski2003estimation, gao2017estimating, paninski2003estimation} under some assumptions, such as i.i.d samples. While \TUPSK does not guarantee sample independence (due to coordination), our experiments (Section~\ref{sec:true-mi-vs-full-join}) show that the estimates converge to true MI when the sample size increases.
Moreover, the high-probability error bounds for the empirical entropy and MI using the MLE estimator proposed in~\cite{wang2019fast} (and subsequently improved in~\cite{chen2021efficient}) apply to sampling without replacement, which is similar to our setting. 
These bounds guarantee that the approximation error (i.e., the difference between an MI estimate computed on a subsample and the MI estimate computed on the full data) reduces in a near square root rate with respect to the subsample size (i.e., the sketch join size, in our case), and allow computing confidence intervals around the estimate that get tighter as the sketch join size approaches the full join size.
While it is unclear if all assumptions in this bound hold for the samples generated by TUPSK, we have also observed this behavior in our experiments.

\section{Experimental Evaluation}
\label{sec:experiments}
To evaluate the efficacy of our proposed sketching methods, we performed experiments using both synthetic and real-world data to answer the following questions:
(Q1)~How accurate are sketches at estimating the true mutual information of attributes obtained after the join? (Section~\ref{sec:experiments-estimation-accuracy-synthetic});
(Q2)~How does join-key distribution affect the MI estimation accuracy? (Sections~\ref{sec:experiments-join-key-effect} and \ref{sec:experiments-distinct-values-synthetic});
(Q3)~How does accuracy vary depending on target and feature data types and, thus, the applied MI estimator?  (Section~\ref{sec:experiments-synthetic-data});
(Q4)~How do sketches behave when estimating MI on real data collections? (Section~\ref{sec:experiments-real-data})

\myparagraph{Mutual Information Estimators}
Many MI estimators have been proposed. We consider a representative set of estimators that are widely used in practice. Unless otherwise noted, we choose estimators based on the data types of variables $X$ and $Y$. When dealing with real data, we consider the following cases: 
(1) If both $X$ and $Y$ have string values (i.e., the discrete-discrete case), we use the maximum likelihood estimator (\textbf{MLE});
(2) If $X$ and $Y$ are numerical variables (e.g., float, integer), we consider the \textbf{MixedKSG} estimator~\cite{gao2017estimating}. This estimator is able to handle not only continuous distributions but also mixtures of discrete and continuous distributions in the same variable, making it flexible for dealing with real data where the distributions are unknown;
(3) When one of the variables is numerical and the other is string (the discrete-continuous case), we use the estimator proposed by Ross in \cite{ross2014mutual} that handles this case, referred to here as \textbf{DC-KSG}.

\myparagraph{Sketching Methods}
We evaluate \CSKext and \TUPSK (Section~\ref{sec:mi-estimation}). 
We also implemented another two-level approach that performs weighted sampling based on key frequencies (using priority sampling~\cite{duffield@jacm2007})
instead of the uniform sampling used in the first level of \CSKext, which we refer to as \PRISK.
Since its results are very similar to those of \CSKext, we omit them for the analysis using synthetic data. 
As baselines, we compare against independent Bernoulli sampling (\INDSK) and a straightforward extension of Correlation Sketches (\CSK)~\cite{santos@sigmod2021}
that estimates MI instead of correlation measures.
Since \CSK 
does not prescribe how to handle repeated join keys,
we use the first value seen associated with a join key (instead of applying an aggregation function that would modify the original values).

\subsection{Synthetic Data Generation}
\label{sec:generating-synthetic-data}
To better understand the behavior of the proposed sketches and answer questions Q1, Q2, and Q3, we used synthetic data to control both the data distribution and join-key dependencies.
We designed a data generation process to create tables given the join key distribution and true MI between $X$ and $Y$ post-join as input.
This is achieved by generating the post-join target $Y$ and feature $X$ by drawing random values from analytic distributions as the join result.
Then we decomposed the join into two separate tables, establishing connections using key $(K_Y)$ and foreign-key ($K_X$) attributes. 
This approach allows us to calculate the true MI after the join, providing a reliable measure to evaluate the effectiveness of our method.

\myparagraphem{Target/Feature Generation}
In the first experiment, we generated random variables $(X, Y)$ using a multinomial distribution $\multinom$, which we refer to as \trinomial. 
This generates three discrete random variables that assume integer values in $\{1, ..., m\}$. Each value represents the number of times that each of the three possible outcomes has been observed in $m$ trials, with each outcome having probabilities $p_1, p_2,\text{ and }(1-p_1-p_2)$.~The number of trials~$m$ is chosen based on the desired number of distinct values in the data. We refer to the first two variables associated with $p_1$ and $p_2$ as $X$ and $Y$, and the third variable is discarded. 

To control the desired level of MI between $X$ and $Y$, we use the following property of the trinomial distribution (see \cite[chapter~11.1]{Georgii+2012}).
The central limit theorem ensures that $\multinom$ converges to a bivariate normal distribution $N(\mu, \sigma)$ with mean $\mu = \langle mp_1, mp_2 \rangle$ and variance $\sigma^2 = m\sqrt{p_i p_j}(\delta_{ij} - \sqrt{p_i p_j})$ as $m \rightarrow \infty$.
Hence, solely for the purpose of selecting model parameters to achieve a desired MI,
we can use the closed-form MI formula from the analogous bivariate normal distribution
to approximate the MI for the trinomial,
which is known to be $-\frac{1}{2} \ln\left( 1 - r^2 \right)$ where $r$ is the Pearson's correlation coefficient of $X$ and $Y$.
Based on this and standard properties of the trinomial distribution, we derived the following algorithm to select the distribution parameters $p_1$ and $p_2$:     
\begin{enumerate}[leftmargin=15pt]
  \item Choose the true mutual information $I_{true} \sim Unif(0,3.5)$, then compute the equivalent correlation $r=$ $\sqrt{1 - \exp{(-2 \cdot I_{true}})}$; note that $I_{true}=3.5$ is equivalent to $r\approx0.999$.
  \item Choose $p_1 \sim Unif(0.15, 0.85)$ (as the approximation works better when $p$ is not near to 0 or 1).
  \item Finally, calculate $p_2$ using the values of $r$ and $p_1$ based on the trinomial variance and closed-form expression for correlation $r = -p_1 p_2 / \left( \sqrt{p_1 (1-p_1)} \sqrt{p_2 (1-p_2)} \right)$. If $p_2$ is not in the desired range (i.e., $[0.15, 0.85]$) then repeat.
\end{enumerate}
The approximation using bivariate normal distribution above was only used to choose the parameters.
To compute the true MI of the distribution, we used the (open-form) entropy formula for the trinomial distribution~\cite{wiki:multinomial}.

As done in~\cite{gao2017estimating}, we also generated a combination of discrete and continuous data for $X$ and $Y$, respectively, which we refer to as \cdu. 
$X$ follows a uniform distribution over the integers $\{0, 1, ..., m-1\}$, while $Y$ is uniformly distributed within the range $[X, X + 2]$ for a given $X$.
The true mutual information between $X$ and $Y$ can be computed as $I(X, Y) = \log(m) - (m-1) \log(2)/m$. Note that here the MI is a function of parameter $m$, which also represents the number of distinct values in $Y$.

\myparagraphem{Distribution Parameters}
For \trinomial, we restrict generated data to having MI $\in [0, 3.5]$ and $m \in \{16, 64, 256, 512, 1024\}$.
Since both $X$ and $Y$ are discrete and have ordered numeric values, it is possible to treat the data as either discrete, mixture, or continuous.
A marginal variable can be made continuous via perturbation, by breaking ties using random Gaussian noise of low magnitude without any significant impact on the MI~\cite{kraskov2004estimating}.
Thus, by doing this in just one of the marginals we can use an estimator for discrete-continuous variable pairs such as the DC-KSG~\cite{ross2014mutual}.
Additionally, MixedKSG~\cite{gao2017estimating} can also be applied to variables with repeated values as it can handle ties naturally based on its formulation.
Hence, to evaluate the impact of these estimators, we consider three representative data type combinations: we use the MLE for discrete-discrete, MixedKSG~\cite{gao2017estimating} for mixture-mixture, and DC-KSG~\cite{ross2014mutual} for discrete-continuous.

For \cdu, we draw $m$ uniformly in the range $[2, 1000]$, which leads to MI values in the range $[0.3, 6.2]$. In this distribution, $Y$ is continuous and $X$ is discrete. Hence, we only report results using MixedKSG~\cite{gao2017estimating} and DC-KSG~\cite{ross2014mutual}, which are able to deal with discrete and continuous distributions seamlessly without any data transformation.

\myparagraphem{Decomposition Into Joinable Tables}
To decompose $(X,Y)$ into tables $\tb{train}$ and $\tb{aug}$ that can be joined to include $X$ and $Y$ as columns, we employ two different methods of generating key and foreign-key columns: \SEQUNIQ for one-to-one joins and \SAMEX for many-to-one joins (which allows us to answer Q2).
\SEQUNIQ provides maximum independence between keys by
generating (sequential) unique join keys in $K_X$, leading to a maximum number of different key values in $K_X$ pointing to the same value in $X$
(e.g., if the value $x$ appears 10 times in $X$ then we have 10 different values in $K_X$ that co-occur with $x$).
This method establishes a one-to-one relationship between attributes $K_Y \in \tb{train}$ and $K_X \in \tb{aug}$.
\SAMEX simulates a strong dependence of join keys by making the value in $K_X$, for each row, equal to the value in $X$.
Hence, we have a single value in $K_X$ for all the occurrences of a value in $X$, establishing a many-to-one relationship. 
Note that \SAMEX is only applicable when $X$ is discrete, as it would create unique join key values for continuous data distributions. Moreover, while the marginal distribution of $X$ (and hence key frequencies in $K_X$) is uniform for \cdu, it is a binomial distribution for \trinomial.
Note also that although these methods represent two contrasting join scenarios, both methods enable table joins that exactly recover $(X, Y)$.


\subsection{Experiments Using Synthetic Data}
\label{sec:experiments-synthetic-data}

\subsubsection{True vs. Estimated MI on Full-Table Joins}
\label{sec:true-mi-vs-full-join}
Before presenting our sketch evaluations, we conducted a preliminary experiment to assess the behavior of different MI estimators.
Our goal is to establish a baseline of the expected behavior of the MI estimators, especially when dealing with real data in Section~\ref{sec:experiments-real-data}, where the true MI is unknown.
For each data distribution, we consider all MI estimators that can be used with the given data types without applying any data transformations as described above: for \trinomial we used MLE, DC-KSG, and MixedKSG; for \cdu we used DC-KSG and Mixed-KSG.
We compare the true MI, calculated via the distribution parameters that were used to generate the data, to MI estimates obtained from the fully-materialized join
containing $N=10$k rows.
%
For both \trinomial and \cdu, the root mean squared error (RMSE) is smaller than 0.07, and the Pearson's correlation coefficient is greater than 0.99.
We omit plots for these results since they are close to a straight line, as expected.
The results demonstrate that MI estimates obtained from the full-table join provide a good approximation for the true MI (computed analytically), 
regardless of the data type assumptions made by each estimator.
Although we can notice some small bias and variance in the range of lower true MI (especially for \trinomial), the overall error is very small in this setting with a large sample size.

\begin{figure}[t]
    \centering
    \includegraphics[width=1\columnwidth]{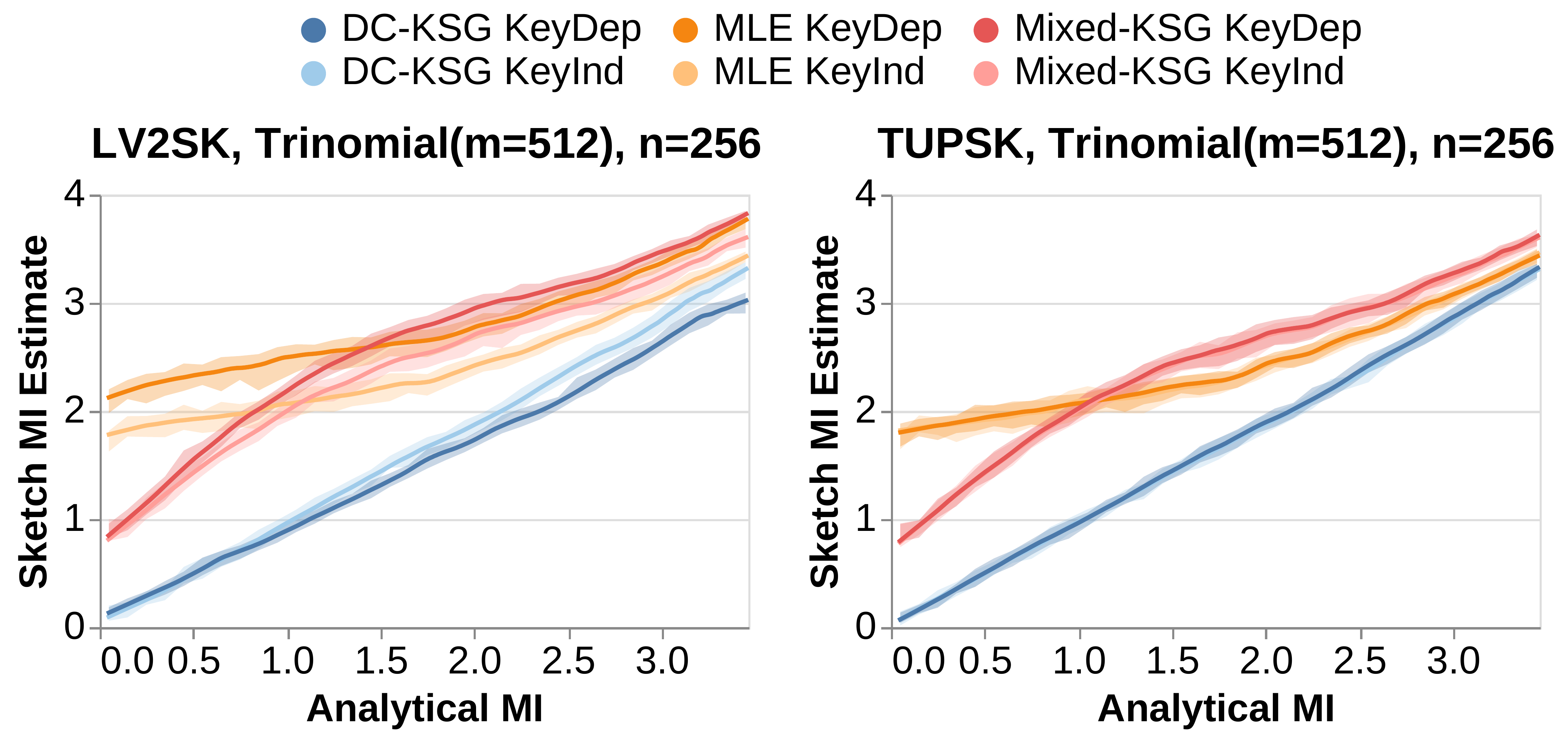}
    \vspace{-.4cm}
    \caption{True MI vs MI estimates computed using sketches of size $n=256$.
    Each plot shows a different 
    method (\CSKext on the left and \TUPSK on the right) and each line shows results for different data types/estimators and join key generation processes.
    \TUPSK is more robust to the join key distribution.
    }
        \vspace{-.55cm}
    \label{fig:mi-dist-vs-sketch}
\end{figure}

\subsubsection{Assessing Sketch Estimation Accuracy} \label{sec:experiments-estimation-accuracy-synthetic}
\autoref{fig:mi-dist-vs-sketch} shows the results for MI estimates for the \trinomial distribution ($m=512$) computed using the proposed sketching methods, \CSKext and  \TUPSK.
In this setting with limited sample size ($n=256$), we see that both the bias and variance of the estimators increase significantly.
Here, the MI is overestimated and the magnitude of the overestimation depends on the type of MI estimator being used: while the bias is highest for MLE estimator (\circleorange,~\circleorangelight) when the true MI is low, MixedKSG (\circlered,~\circleredlight)  reaches a peak bias around the mid-range MI values.
\hide{
MixedKSG: (\circlered,~\circleredlight) 
DC-KSG:   (\circleblue,~\circlebluelight)
MLE:      (\circleorange,~\circleorangelight)
}

While the bias and variance are also influenced by other factors (as discussed below), this result underscores the significance of selecting the appropriate estimator for the data type at hand. For example, while it may be simpler to use an MLE estimator with discrete ordered data (or, e.g., with binned continuous data), using a $k$-Nearest Neighbors approach may lead to a smaller bias.

Furthermore, this result suggests that comparing estimates of columns with different data types, which require distinct estimators, may not yield meaningful results due to the distinct bias and variance properties of different estimators. While the results of these estimators converge to the true MI when the sample size is large enough, as shown in Section~\ref{sec:true-mi-vs-full-join}, the approximation accuracy depends on many factors such as the data distribution, the sample size, and the underlying true MI.
When these are unknown, it becomes challenging to determine whether such comparisons are meaningful. In Section~\ref{sec:experiments-real-data}, we further discuss this issue based on our results on data sourced from real-world open data repositories.

\subsubsection{Effect of the Join Key Distribution} \label{sec:experiments-join-key-effect}
In Section~\ref{sec:mi-estimation}, we described two different methods to select samples to include in the sketch. In Figure~\ref{fig:mi-dist-vs-sketch}, we can visualize how the join-key distribution affects the MI estimation accuracy of these methods. 
Specifically, we can compare the estimation difference caused by \SEQUNIQ vs \SAMEX in a given estimator. For instance, consider the \CSKext method in Figure~\ref{fig:mi-dist-vs-sketch}(left). When we compare the lines that represent the MLE estimator (\circleorange,~\circleorangelight), we note that the bias from \SAMEX (\circleorange) is larger than that from \SEQUNIQ (\circleorangelight).
Similarly, \SAMEX leads to increased bias for the MixedKSG estimator (\circlered) but, for the DC-KSG estimator, \SAMEX leads to a small downward bias (\circleblue).

Differently from \CSKext, \TUPSK is not as affected by the join-key distribution. We can see in Figure~\ref{fig:mi-dist-vs-sketch}(right) that \TUPSK is able to attain the same performance regardless of the join-key distribution.  This is because the \TUPSK sampling scheme reduces the dependence on the join keys by hashing on the tuple $\langle k, j \rangle$, which leads each row being sampled with uniform probability (given that each tuple $\langle k, j \rangle$ is unique in table $\tb{train}$). \CSKext on the other hand, samples entries non-uniformly and introduces additional bias due to its dependence on the key frequency distribution and the existing key-target correlation in the \SAMEX distribution.

\begin{figure}[t]
    \centering
    \includegraphics[width=1\columnwidth]{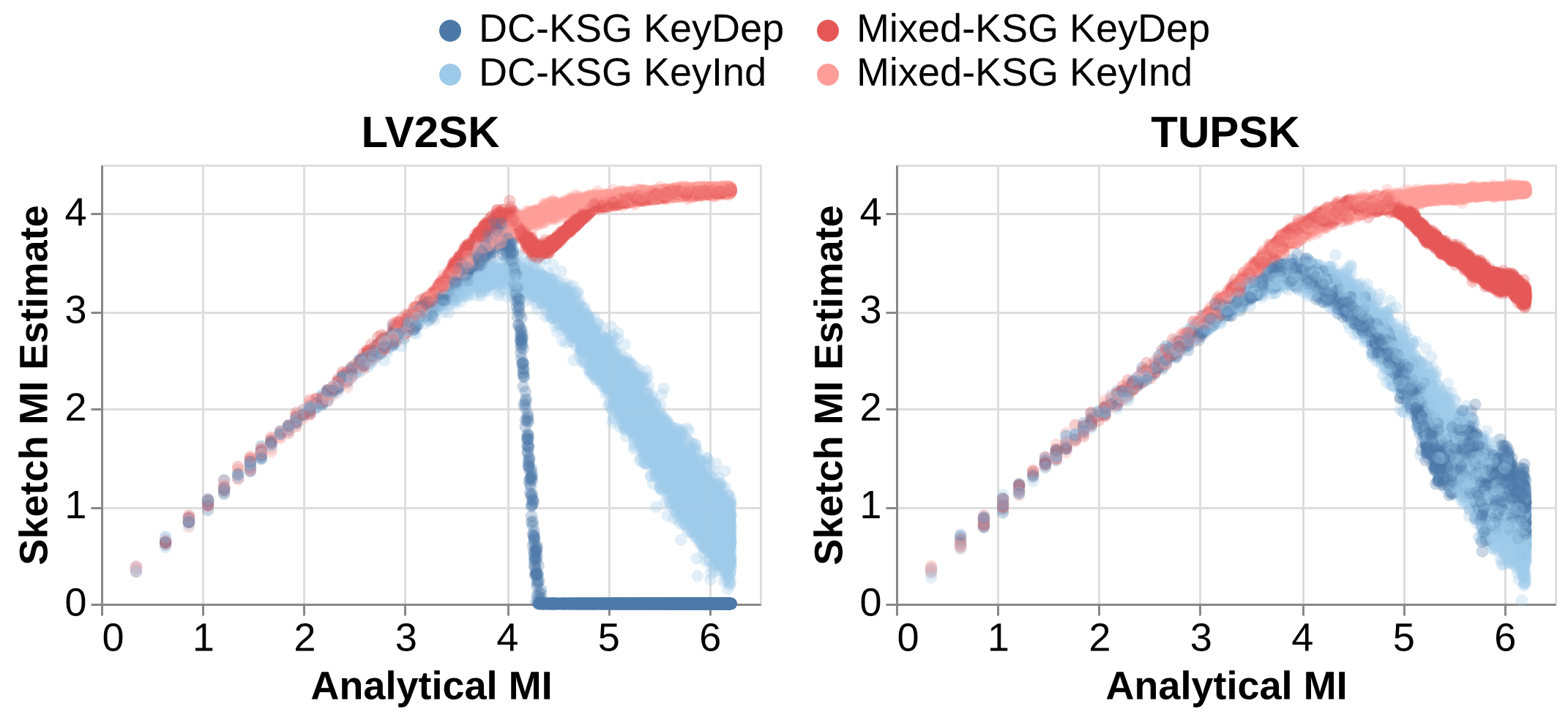}
    \vspace{-.4cm}
    \caption{True MI vs MI estimates computed using sketches of size $n=256$ for \cdu. Each plot shows a different sketching method while each line shows results for different data types/estimators and join key generation processes.
    }
    \vspace{-.5cm}
    \label{fig:cdunif-mi-dist-vs-sketch}
\end{figure}

\begin{figure*}[t]
    \centering
    \includegraphics[width=1\textwidth]{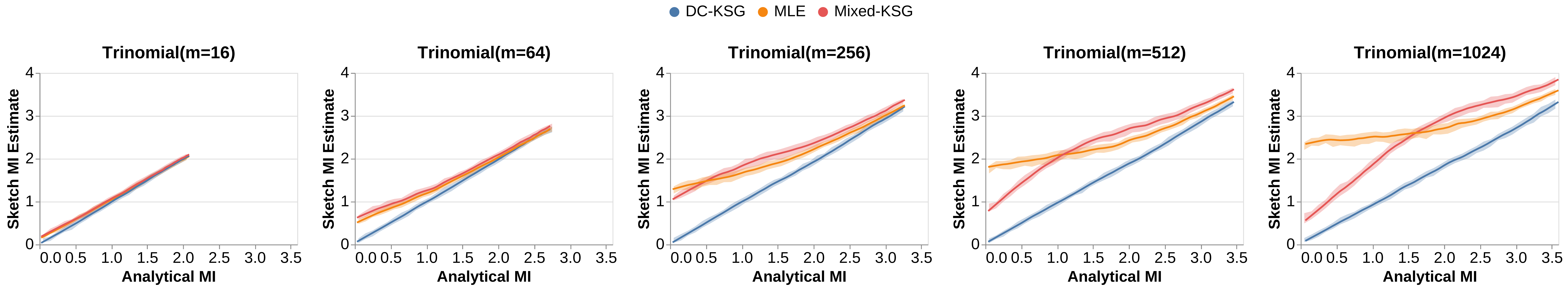}
    \vspace{-.45cm}
    \caption{Sketch MI estimate versus the true MI computed using distribution parameters. Sketch size is $n=256$ for all plots.
    }
    \vspace{-.35cm}
    \label{fig:loess-mi_dist-mi_est-_multinomial-m=16-to-1024}
\end{figure*}

\subsubsection{Effect of Distinct Values}   \label{sec:experiments-distinct-values-synthetic}
To assess the impact of the number of distinct values in the distribution on the MI estimation accuracy, we vary the parameter $m$ of \trinomial and \cdu distributions while keeping the desired sketch size $n=256$ constant. This means that the ratio $m/n$ increases making it increasingly harder to estimate the MI, e.g., when $X \sim Uniform$ and $m/n=1$ we expect to have only 1 sample to estimate the probability mass $p(x)$ of a given value $x \in X$.
For \cdu, $m=256$ is equivalent to $I(X,Y) \approx 4.85$. As Figure~\ref{fig:cdunif-mi-dist-vs-sketch} shows, the MI estimators break down when $I(X,Y)$ approaches $4.85$ for the \cdu distribution. For \CSKext, the DC-KSG estimator completely breaks down even earlier, around $I(X,Y) \approx 4.25$. In contrast, \TUPSK degrades more gracefully as $I(X,Y)$ increases.

Figure~\ref{fig:loess-mi_dist-mi_est-_multinomial-m=16-to-1024} shows the impact of increasing $m$ for the \trinomial distribution. Here, the marginal distributions of $X$ and $Y$ are non-uniform (binomial) distributions (unlike \cdu which has uniform marginals). The plots clearly show that increasing $m$ leads to increased bias for estimators that handle discrete distributions such as MLE (\circleorange) and MixedKSG (\circlered). Although the estimators do not completely break down here, we can see that the bias for the MLE estimator (\circleorange) is so large when $m=1024$ that all estimates are considered to have a high MI in the small range $[2.5, 3.5]$.

Note that the maximum true MI value for low values of $m$ is smaller than for large $m$. 
This is due to our data generation process (Section~\ref{sec:generating-synthetic-data}),
which relies on the central limit theorem and the bivariate normal distribution to approximate the MI. 
This, however, does not affect our results since we use the 
exact MI formula to compute the analytical MI in~Fig.~\ref{fig:loess-mi_dist-mi_est-_multinomial-m=16-to-1024}.

\subsubsection{Comparison to Other Baselines} 
\label{sec:experiments-other-baselines}

In Table \ref{tab:results-synthetic-data}, we report the average sketch join size and mean squared error (MSE) for all sketches, including the additional baselines: independent sampling (\INDSK) and correlation sketches (\CSK). Results are computed for sketches of size $n=256$ and include tables with different join key distributions (\SAMEX, \SEQUNIQ) and different distribution parameters ($m$).
The results demonstrate that \INDSK has difficulty matching join keys, resulting in a smaller join size than coordinated sampling approaches, which leads to large MSE. Coordinated sampling methods achieve significantly larger join sizes, making them more effective strategies. Among them, \TUPSK achieves the best MSE, which is due not only to the larger number of samples recovered by the sketch join but also to unbiased samples.
The average join size of the two-level sampling sketches is highly sensitive to the join key distribution: when keys are unique (as in \SEQUNIQ), it behaves as \TUPSK, and a sketch of size $n$ yields $n$ join samples. However, when there are repeated keys, it may lead to either more or fewer samples than $n$. In our experiments, the average join size increases as $m$ increases.

\begin{table}[]
    \centering
    \begin{small}
    \setlength{\tabcolsep}{4pt}
    \begin{tabular}{ccccc}
    \toprule
    \textbf{Dataset} & \textbf{Sketch} &  \textbf{Avg. Sketch Join Size} &  \textbf{\%}  & \textbf{MSE} \\
    \midrule
    \multirow[c]{5}{*}{\cdu} & CSK & 194.2 & 75.87 & 4.56 \\
     & INDSK & 107.9 & 42.16 & 9.57 \\
     & LV2SK & 232.9 & 90.99 & 2.94 \\
     & PRISK & 232.9 & 90.99 & 2.94 \\
     & TUPSK & 256.0 & \bfseries 100.00 & \bfseries 0.77 \\ \midrule
    \multirow[c]{5}{*}{\trinomial} & CSK & 155.2 & 60.62 & 1.37 \\
     & INDSK & 133.7 & 52.22 & 1.19 \\
     & LV2SK & 255.9 & 99.94 & 0.32 \\
     & PRISK & 255.9 & 99.94 & 0.32 \\
     & TUPSK & 256.0 & \bfseries 100.00 & \bfseries 0.22 \\
    \bottomrule
    \end{tabular}
    \end{small}

    \caption{
    Comparison of MI estimate versus the true MI using sketches of size $n=256$. The ``\%'' column is the percentage of ``Avg. Sketch Join Size'' relative to sketch size~$n$. 
    }
    \vspace{-.5cm}
    \label{tab:results-synthetic-data}
\end{table}

\vspace{-.15cm}
\subsection{Experiments Using Real Data}
\label{sec:experiments-real-data}

We now evaluate the behavior of our sketches on real-world data collected from two different open-data portals: the World Bank’s Finance (WBF)~\cite{wbopendata-finance} and the NYC Open Data~(NYC)~\cite{nycopendata}.
Our experimental data consists of snapshots of these repositories collected in September 2019 using Socrata’s REST API~\cite{socrata}. 

From these collections, we sample pairs of tables $\tb{train}$ and $\tb{aug}$ as follows.
For each table $t$ in a data repository, we first create the set $\mathcal{P}_t$ of two-column tables, denoted as  $\tb{A}[K_A, A]$, comprised of all pairs of join-key and data attributes $\langle K_A, A \rangle$ from $\mathcal{P}_t$ such that $K_A$ is a string attribute and $A$ contains either strings or numbers (i.e., ints, longs, floats, or doubles).\footnote{We used the Tablesaw library~\cite{tablesaw} to perform type inference.}
Let $\mathcal{C} =  \bigcup_t \mathcal{T}_t$ be the set of all two-column tables in the repository. We then draw a uniform sample of the set of pairwise combinations $\mathcal{PC} = \{(\tb{i}, \tb{j}) \mid \tb{i}, \tb{j} \in \mathcal{C}\}$ and use the tables in these pairs as $\tb{train}$ and $\tb{aug}$.
The final sample includes 36k table pairs for the WBF collection and 59k pairs for the NYC collection.
The average domain size of join attributes for the left and right tables are approximately 3.1k and 3.5k for WBF, respectively, and 11.2k and 1k for NYC, respectively.
Finally, the average full join size is 34k for WBF and 8.5k for NYC.

Given that it is not possible to know the true distribution of the data in these tables, we use the MI estimated over the full data as a proxy for the true MI. As shown in Section~\ref{sec:true-mi-vs-full-join}, the full join provides a good approximation of the true MI when the join size is large. Hence, from now on we compare the sketch estimates to the full-join estimates. Even though the full join may not always reflect the true MI, it is the only option available in many practical scenarios~\cite{wang2019fast}.

\subsubsection{Approximation Accuracy} Table~\ref{tab:sketch_size} summarizes the results for each sketching method for the two dataset collections. The results are computed using sketches with $n=1024$. To discard meaningless estimates, we only include estimates computed on sketch join size greater than 100.
First, we confirm that \CSKext, which may use a higher storage size for a given budget $n$ (see Section~\ref{sec:csk-two-level}), tends to generate a larger average join size.
However, despite using less storage, \TUPSK outperforms \CSKext in terms of estimation accuracy measured by the mean squared error (MSE) metric.

We use Spearman's correlation, a rank-based measure, to quantify how well the ranking obtained using MI estimates computed from sketches approximates the ranking of MI estimates computed over the full tables. 
We can see that Spearman's correlation for \TUPSK is the strongest.  
This is significant since for automatic data augmentation it is important to rank features based on their importance.
This result confirms that \TUPSK is able to generate higher quality samples than its competitors.

\begin{table}[t]
    \vspace{.2cm}
    \centering
    \small
    \begin{tabular}{ccccc}
    \toprule
    \textbf{Dataset} & \textbf{Sketch} &  \textbf{Avg. Join Size} &  \textbf{Spearman's R}  & \textbf{MSE} \\
    \midrule
    \multirow[c]{3}{*}{NYC} & LV2SK & 230.9 & 0.81 & 1.41 \\
     & PRISK & 231.1 & 0.79 & 1.36 \\
     & TUPSK & 185.3 & \bfseries 0.86 & \bfseries 0.93 \\ \midrule
    \multirow[c]{3}{*}{WBF} & LV2SK & 231.2 & 0.40 & 1.75 \\
     & PRISK & 226.6 & 0.40 & 1.76 \\
     & TUPSK & 194.9 & \bfseries 0.45 & \bfseries 1.46 \\
    \bottomrule
    \end{tabular}
    
    \caption{Comparison of MI estimate using different sketching strategies versus the full join. While \CSKext can theoretically have a sketch size twice as large as \TUPSK, in practice their sketch join sizes is similar. Even with this disadvantage, \TUPSK outperforms \CSKext in estimation accuracy (stronger Spearman's R correlation) using less storage.}
    \vspace{-.55cm}
    \label{tab:sketch_size}
\end{table}

\subsubsection{Effect of Sketch Join Size} 
\begin{figure*}[t]
    \centering
    \includegraphics[width=.99\textwidth]{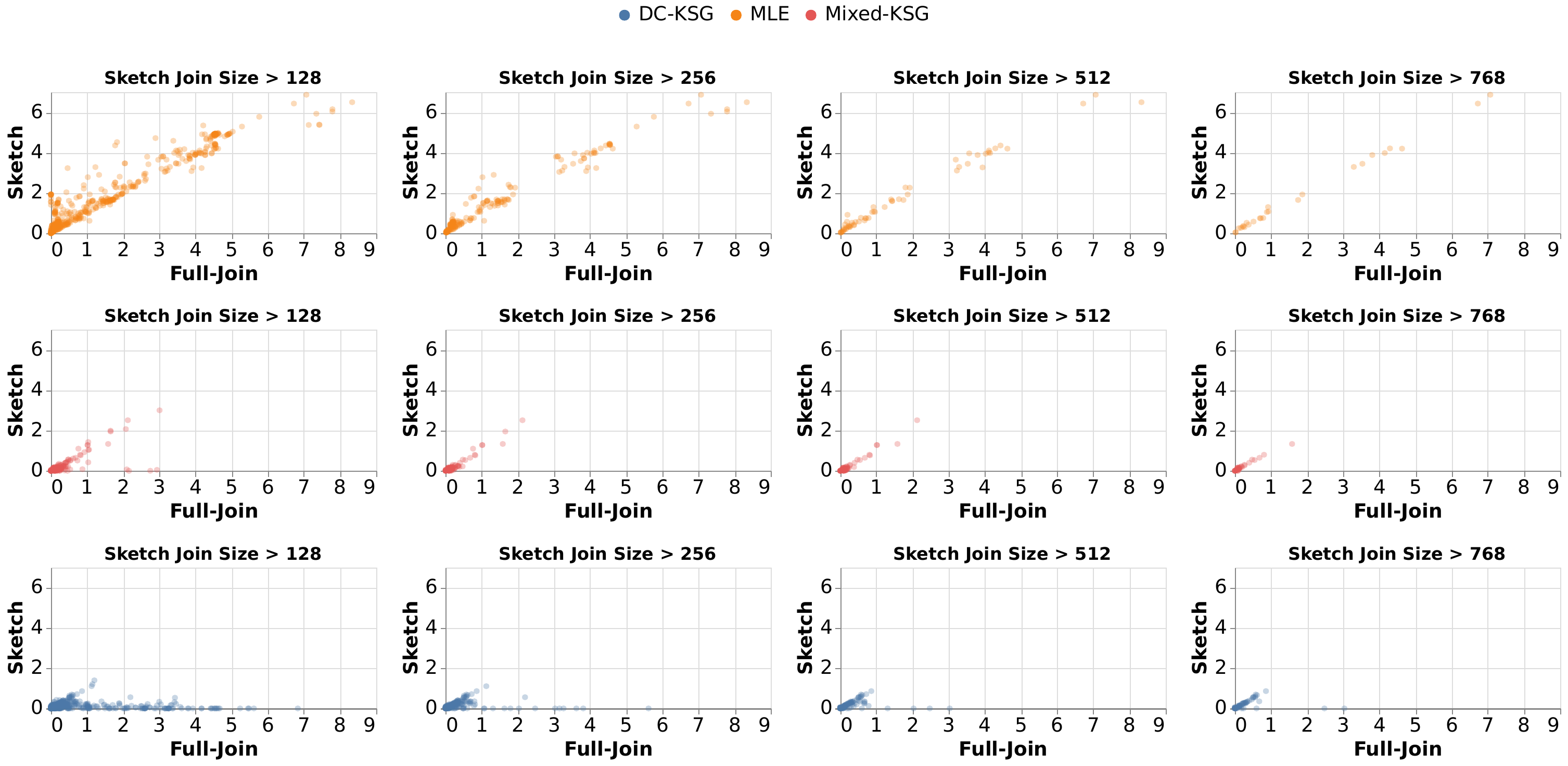}
    \vspace{-.25cm}
    \caption{Sketch MI estimate versus the MI estimate computed using the full join output for tables from the WBF collection. Sketches are created using \TUPSK with size $n=1024$ for all plots.}
    \vspace{-.5cm}
    \label{fig:wbf_mi-sketch_vs_mi-fulljoin}
\end{figure*}

In Figure \ref{fig:wbf_mi-sketch_vs_mi-fulljoin}, we break down the results by data types (and hence, MI estimators) and sketch join size. A larger sketch join size indicates that the tables are more joinable (i.e., have a larger overlap) and that the MI estimator is given a larger number of samples. Here, we can observe a behavior similar to what we observed with synthetic data. In particular, we note that when the sample size is small, (1) the MLE estimator (\circleorange) tends to overestimate the MI, and (2) the KSG-type estimators (\circlered,~\circleblue) tend to break down and generate estimates close to zero.

\subsubsection{Comparing MI Estimators} 
Another notable difference is the magnitude of the MI estimates generated by different estimators: the MLE estimator computes MI values that are significantly larger than the ones generated by KSG-based estimators: while MLE estimates reach the range $[4, 6]$, KSG-based estimates are never larger than 2. Although we cannot confirm whether this is an artifact of the estimator limitations or if numerical data indeed leads to smaller MI values, this result suggests that comparing MI estimates from different estimators may not be reasonable. For example, when ranking attributes for data discovery, it might be preferable to produce separate rankings of different MI estimators and then compare the utility of top-ranked attributes using a downstream (task-specific) evaluation measure (e.g., the increase in accuracy of an ML model computed on the labels).

\subsection{Performance Evaluation} 
Due to space constraints, we omit a detailed runtime evaluation since the efficiency of the proposed sketches is similar to others evaluated in previous work~\cite{santos@sigmod2021, santos@icde2022}.
For completeness, we provide exemplar numbers for sketch size $n=256$. 
As the table size grows from $N=5k$ to $N=20k$, the full join size time increases from 0.35ms to 2.1ms, whereas the sketch join time grows from 0.03ms to 0.18ms.
Similarly, while MI estimation time increases from 2.2ms to 10.7ms, the sketch is approximately constant and took only 0.1ms.

\section{Related Work}
\label{sec:related-work}

To the best of our knowledge, no prior work has addressed the problem of estimating mutual information (MI) over joins for relational data augmentation using sketches.
In what follows, we examine related prior research on data discovery systems, feature selection, MI estimation, and join sampling.

\paragraph{Data Discovery}
The problem of finding related datasets (via unions and joins) on the Web and in data lakes, to unlock the utility of a provided table has been studied since~\cite{Sarma2012FindingRT}.
Applications include decision support, data mining, ML model improvement, and causality analysis, and have resulted in systems including Aurum~\cite{fernandez@icde2018}, ARDA~\cite{chepurko2020arda} and Auctus~\cite{auctus@vldb2021}.
Some related work focuses on methods that utilize sketches and Locality-Sensitive Hashing (LSH) indexes to efficiently discover joinable tables~\cite{zhu@vldb2016-lsh-ensemble, fernandez@icde2019}. Others address the problem of finding tables that are both joinable with the input query table and contain correlated columns, employing different techniques to identify these relationships~\cite{kumar2016join, santos@sigmod2021, santos@icde2022, esmailoghli2022mate, becktepe2023demonstrating}.
MI has been used for discovering functional dependencies (FDs), often between columns within the same table~\cite{pennerath2020discovering,mandros2020discovering,mandros2019discovering}.
FDs are not symmetric, unlike MI, so while MI can help discover FDs, the reverse is not always true.
Finally, our method is complementary to the work
of Kumar et. al. \cite{kumar2016join}, where they propose conservative decision rules to predict when the features obtained through a join can improve models.

\paragraph{Feature Selection}
There are three main classes of feature selection methods (see~\cite{vergara2014review} for a survey): filter methods, which start from a join over all tables and use a lightweight proxy such as correlation to remove features; wrapper methods, which incrementally choose features based on the (more expensive) downstream task either by joining one
table/feature at a time (forward selection) or removing one at a time from a join over all tables; and embedding methods, which use a proxy to select multiple features at a time and then measure using the downstream task.
This work is related to filter methods since it enables the estimation of MI, which is a proxy measure commonly used in several feature selection algorithms~\cite{vergara2014review, brown2012conditional}.

\paragraph{Mutual Information Estimation}
When the number of available observations is large, computing the MI can be expensive.
Various papers have considered efficient approximation algorithms, with confidence intervals, for entropy estimation via subsampling~\cite{wang2019fast, chen2021efficient}, which can be extended to MI.
Some approaches considered MI estimation over data streams \cite{10.5555/1347082.1347163,10.1145/2791347.2791348,10.1145/3019612.3019669}.
Ferdosi et. al. \cite{pmlr-v108-ferdosi20a} show how to approximately find a pair of columns having the largest mutual information in sub-quadratic time, however, they assume binary-valued attributes and that table joins are materialized.
While our prior work on Correlation Sketches~\cite{santos@sigmod2021} examined the problem of sketches to avoid a full join, we are not aware of any work addressing the problem of estimating MI while avoiding the cost of materializing the entire join.
There is also an extensive body of research on different estimators for MI, some of which we cover in Section~\ref{sec:background}. Additionally, recent work has shown that no estimator can \textit{guarantee} an accurate estimate of mutual information without making strong assumptions on the population distribution~\cite{mcallester2020formal}. While previous work had provided intractability results for specific estimators, such as KSG \cite{gao2015efficient}, this result is universal to all MI estimators.

\paragraph{Join Sampling}
The problem of creating samples of table join results 
has been extensively studied~\cite{huang2019joins, chen@sigmod17, vengerov@vldb2015, estan@icde2006endbiased}. We use ideas similar to the ones employed by these algorithms, such as sample coordination and two-level sampling. Unlike our sketches, however, these algorithms lead to variable sample sizes that depend on the input size and data distribution. Moreover, they usually employ unequal probability sampling that requires adjusting the estimators to remove bias. In contrast, we seek sampling strategies that lead to uniform probabilities that allow us to use existing MI estimators.

\vspace{-.15cm}
\section{Conclusion}
\label{sec:conclusion}
In this paper, we introduced a new method for estimating mutual information (MI) using sketches.
Our proposed method addresses the challenges of estimating MI over joins with non-unique keys and efficiently approximates the MI without the need to materialize the full join.
Our experiments demonstrate the effectiveness of our sketches in approximating the true MI values. Additionally, we have shown how data types and estimators impact the accuracy of MI estimates, emphasizing the need for careful consideration when comparing MI estimates from different data types.
Our results show that some estimators underestimate while others overestimate, there is thus a need to carefully choose an appropriate estimator based on application requirements.
E.g., while MLE may offer high recall, estimators based on Laplace smoothing [34] may be more appropriate for controlling false discoveries. Exploring this trade-off further is a promising avenue for future work.

\myparagraph{Acknowledgments}
This work was partially supported by the DARPA D3M program, NSF award ISS-2106888, and a Google Research Collabs grant. Any opinions expressed in this material are those of the authors and do not necessarily reflect the views of the funding organizations.

\balance
\bibliographystyle{IEEEtran}
\bibliography{paper}

\end{document}